\documentclass[useAMS,usenatbib]{mn2e}

\usepackage{epsfig}

\title[Oscillations of massive stars]{Linear analyses for the stability of radial and nonradial oscillations of massive stars}
\author[H. Saio]
{Hideyuki Saio\thanks{E-mail:saio@astr.tohoku.ac.jp}\\
Astronomical Institute, Graduate School of Science, Tohoku University,
 Sendai, Miyagi 980-8578, Japan}

\date{}

\pagerange{\pageref{firstpage}--\pageref{lastpage}} \pubyear{2010}

\begin{document}

\label{firstpage}

\maketitle

\begin{abstract}
In order to understand the periodic and semi-periodic variations
of luminous O- B- A-type stars, linear nonadiabatic stability analyses 
for radial and nonradial oscillations have been performed for massive 
evolutionary models ($8M_\odot - 90M_\odot$).  
In addition to radial and nonradial oscillations excited by  
the kappa-mechanism and strange-mode instability,
we discuss the importance of 
low-degree oscillatory convection (nonadiabatic g$^-$) modes.
Although their kinetic energy is largely confined to the convection 
zone generated by the Fe opacity peak near $2\times10^5$K,   
the amplitude can emerge to the photosphere and should be observable 
in a certain effective temperature range. 
They have periods longer than those of the radial strange modes so that 
they seem to be responsible for some of the long-period microvariations 
of LBVs (S Dor variables) and $\alpha$ Cyg variables.
Moreover, monotonously unstable radial modes are found in some models whose
initial masses are greater than or equal to $60M_\odot$ with $Z=0.02$. 
The monotonous instability probably corresponds to the presence of an optically
thick wind. The instability boundary roughly coincides with the
Humphreys-Davidson limit.  
\end{abstract}


\section{Introduction}
It is known that various instabilities occur in very luminous stars. 
Most luminous stars called Luminous Blue Variables (LBVs or S Dor variables) 
show major events (SD(S Dor)-eruptions and SD-phases) with which mass loss 
from the star is thought to be enhanced greatly 
\citep[see e.g.,][for reviews]{hd94,vanG01}. Their distribution
on the HR diagram seems to be bounded by the Humphreys-Davidson 
\citep[HD;][]{hd79} limit.
\citet{ki93} found that the instability boundary for the radial strange modes 
roughly coincides with the HD limit.  They  proposed that 
the strange mode instability would yield a
strong mass loss causing the HD limit.
Several nonlinear analyses for radial strange modes have been
performed \citep{do00,ch04,gr05} to generate pulsation driven mass-loss,
but the results seem to remain inconclusive.
(In this context, it is interesting to note that \citet{ae10} 
found a luminous B star 
to change its mass-loss rate on a timescale of the period of photometric 
and spectroscopic variation, suggesting a correlation between mass-loss rate 
and pulsation.)

It is fair to mention that various instabilities other than strange mode
instability 
are also proposed for the cause of eruptive mass loss 
\citep[see e.g.,][for a review]{vanG01}.
In the present paper (\S\ref{sec:secular}) we will discuss the presence of
a monotonously unstable mode in models more luminous
than the HD limit.

In addition to the major variations (SD-eruptions and SD-phases) 
on timescales of 
years and decades, LBVs show quasi-periodic microvariations on
timescales of weeks to months.
Those variations occur also in non-LBV luminous OBA supergiants. 
They are loosely called $\alpha$ Cyg variables, and
the variations are thought to be caused by stellar oscillations.
\citet{dz05} and \citet{fa10} claimed these microvariations
to be caused by radial strange mode pulsations.

\citet{Lam98} argued, however, that periods of some of the microvariations
in LBVs are much longer than the periods of strange modes, and that
the positions of some of those variables on the HR diagram contradict
the strange mode instability region.
They concluded microvariations in LBVs to be consistent with nonradial g-modes.

In the present paper, we find those longer period variations 
can be interpreted by oscillatory convection (nonadiabatic g$^-$) modes 
associated with the envelope convection
zone generated by the Fe-opacity peak around $T\sim 2\times10^5$K.
The linear convection modes, which are dynamically (monotonously) 
unstable in the adiabatic
analysis (g$^-$-modes), become overstable (or oscillatory) if nonadiabatic
effects are included as found by \citet{sh81} for high-degree ($\ell \ga 10$)
modes. 
These modes have not been paid much attention to before, because
they are not expected to be observed due to 
the high $\ell$ values and their amplitude being 
confined to the convective zone.
The present paper will show some low-degree ($\ell \le 2$) overstable 
convection modes having periods longer
than those of radial strange modes can emerge to the stellar surface
and hence be observable. 

Before the emergence of the OPAL opacity \citep{ro92}, 
OB-type stars were considered to have mostly radiative envelopes.
But in reality, the Fe-opacity peak at $T\sim2\times10^5$K in the OPAL opacity
is strong enough to produce a convective zone of considerable
thickness in the envelopes of OBA stars.
\citet{ca09} argued the importance of the sub-photospheric convection 
in massive stars for various photospheric velocity fields including pulsations 
and large microturbulence needed in spectroscopic analyses
of OB stars.
Furthermore, \citet{de10} found evidence of solar-like oscillations
in a O-type star that indicates stochastic excitation of oscillations
by turbulent convection to work even in  hot stars.

There are also less luminous B supergiants which show periodic
microvariations with relatively short periods of one to a few days
\citep{wa98,Lef07}. 
Some of these variations  can be interpreted as supergiant extension 
of SPB stars (SPBsg),
which is possible because g-mode oscillations being reflected at a convection
zone above a hydrogen burning shell so that strong 
dissipation in the core is suppressed \citep{sa06,afg09,go09}.

The present paper discusses the stability of radial and nonradial oscillations
in massive main-sequence and post-main-sequence
models, and
compares periods of excited modes with observed periods of
microvariations in supergiants.

\section{Models and assumptions}
In order to obtain unperturbed models for stability analyses 
covering the upper part of the HR diagram, 
evolutionary models in a mass range of $90\ge M_i/M_\odot \ge 8$ 
were calculated from ZAMS to $\log T_{\rm eff} \approx 3.7$,
using a code based on the Henyey method with OPAL opacity tables \citep{opal},
where $M_i$ means initial mass.
The mixing length is assumed to be 1.5 pressure scale heights 
(unless otherwise stated).
Wind mass loss based on \citet{vink01} is included
for the models with initial mass of $M_i\ge 30M_\odot$. 
Stellar rotation and core overshooting are disregarded for simplicity.
Two sets of chemical compositions are adopted in systematic calculations;
$(X,Z)=(0.70,0.02)$ and $(0.716,004)$. An additional
model sequence was calculated for $M_i=70M_\odot$  with the composition 
$(X,Z) = (0.71,0.01)$ 
to clarify the effect of heavy element abundances
on the stability of oscillations.

\begin{table}
\caption{
Parameters for selected models at ZAMS, TAMS and at $\log T_{\rm eff}=4.0$ 
}
\label{modelparam}
\begin{tabular}{@{}lccccc}
\hline
$M_i/M_\odot$ & $90$ & $70$ & $50$ & $40$ & $30$ \\
\hline
&& \multicolumn{2}{c}{$Z = 0.02$} \\
\hline
ZAMS\\
~~$\log L$ & 6.034 & 5.841 & 5.561 & 5.358 & 5.074\\
~~$\log T_{\rm eff}$ &  4.690 & 4.682 & 4.659 & 4.637 &4.602 \\
\hline
TAMS\\
~~$t(10^6{\rm yr})$ & 2.997 & 3.233 & 3.715 & 4.214 & 5.132\\
~~$M/M_\odot$ & 51.40 & 47.54 & 39.66 & 34.20 & 27.60\\
~~$\log L$ & 6.095 & 5.953 & 5.725 & 5.555 & 5.318\\
~~$\log T_{\rm eff}$ & 4.226 & 4.173 &  4.297 & 4.383 & 4.424\\
\hline
$log T_{\rm eff}= 4.0$ \\
~~$t(10^6{\rm yr})$ &  3.042 & 3.275 & 3.760 & 4.287 & 5.266\\
~~$M/M_\odot$ & 50.37 & 47.09 & 39.39 & 33.93 & 27.41\\
~~$\log L$ & 6.138 & 5.984 & 5.763 & 5.622 & 5.422\\
\hline
\hline
&& \multicolumn{2}{c}{$Z = 0.004$} \\
\hline
ZAMS\\
~~$\log L$ & 6.037 & 5.845 & 5.566 & 5.366 & 5.087\\
~~$\log T_{\rm eff}$ & 4.744 & 4.726 & 4.698 & 4.675 & 4.640 \\
\hline
TAMS\\
~~$t(10^6{\rm yr})$ & 2.895 & 3.181 & 3.747 & 4.389 & 5.637\\
~~$M/M_\odot$ & 80.05 & 63.68 & 46.80 & 38.05 & 29.07\\
~~$\log L$ & 6.207 & 6.039 & 5.800 & 5.632 & 5.406\\
~~$\log T_{\rm eff}$ & 4.428 & 4.4891 & 4.517 & 4.514 & 4.492 \\
\hline
$log T_{\rm eff}= 4.0$ \\
~~$t(10^6{\rm yr})$ &   2.925 & 3.273 & 3.845 & 4.556 & 6.007\\
~~$M/M_\odot$ & 79.82 & 63.47 & 46.68 & 37.92 & 28.91\\
~~$\log L$ & 6.246 &  6.100 & 5.894 & 5.748 & 5.548\\
\hline
\end{tabular}
\end{table}

Table~\ref{modelparam} gives stellar parameters at ZAMS, TAMS 
(the end of main-sequence), and at $\log T_{\rm eff}=4.0$ for selected
model sequences.
It is apparent that significant mass is lost during evolution
for $M_i \ga 50 M_\odot$ with $Z=0.02$.
For the low metallicity ($Z=0.004$) mass loss is considerably weaker.

Nonadiabatic radial and nonradial pulsation analyses were performed 
using the formulae 
described in \citet{sa83} and \citet{sa80}, respectively. 
The Lagrangian perturbation of the divergence of the convective flux 
is neglected.
The temporal variation of radial and
nonradial pulsations is expressed as $\exp(i\sigma t)$ with
$\sigma$ being eigenfrequency whose imaginary part determines
the stability;i.e, the pulsation is excited if the imaginary part 
of $\sigma$ is negative. In numerical calculations the
frequency is normalized as $\omega \equiv \sigma/\sqrt{GM/R^3}$ with
the gravitational constant $G$, stellar mass $M$, and radius $R$.

Although wind mass loss is included in evolutionary models for 
$M_i\ge 30M_ \odot$, for oscillation analyses a reflective mechanical 
outer boundary condition ($\delta P_{\rm gas} \rightarrow 0$) 
is employed because theory for the boundary condition in the presence
of wind has not yet been developed.

\section{Excited modes}

\begin{figure}
\epsfig{file=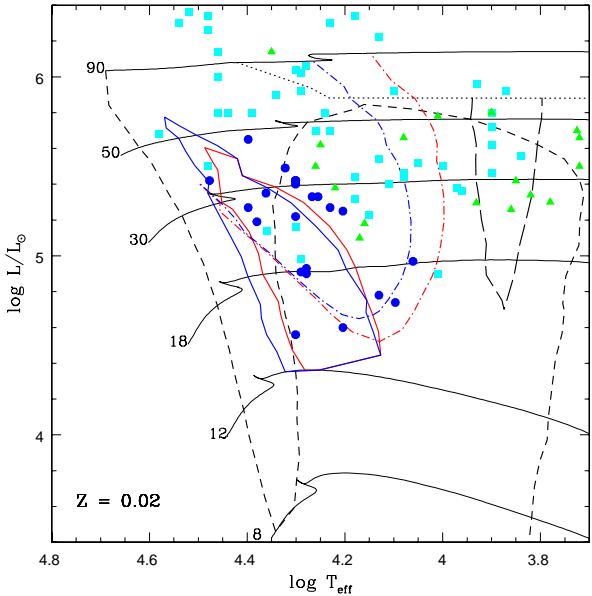,width=0.49\textwidth}
\caption{Instability boundaries of various modes with selected evolutionary
tracks. Black short-dashed and long-dashed lines are for low-order and
high-oder (radial and nonradial) p-modes, respectively. 
Dotted line is the boundary above which monotonously unstable modes 
present.
Red and blue solid lines are for
nonradial g-modes with $\ell=1$ and $\ell=2$, respectively.
Dash-dotted lines are for oscillatory convection modes (red line for $\ell=1$,
blue line for $\ell=2$) indicating
the boundaries of the region where 
the ratio of the photospheric to the maximum amplitude in the interior
is larger than 0.2.  
The number written at the start of each evolutionary track indicates the 
initial mass $M_i$ in solar mass units. Wind mass loss included 
for the evolutionary models with $M_i \ge 30 M_\odot$ based on \citet{vink01}. 
Also shown are positions of super-giant variable stars; 
variable B-type supergiants \citep[blue dots][]{Lef07}, 
S Dor variables \citep[cyan squares][]{vanG01}, 
and $\alpha$ Cyg variables \citep[red triangles][]{vanL98,vanG02}. 
}
\label{fig:hrd}
\end{figure}

Fig.~\ref{fig:hrd} shows various instability boundaries (visibility
boundaries for oscillatory convection modes) obtained in this paper
and selected evolutionary tracks for a standard composition
of $(X,Z) = (0.7,0.02)$. For comparison, positions of periodic
and semi-periodic supergiant variables are plotted from various sources
of the observational data.
Interestingly, most of those variable supergiants reside in the unstable
regions. 

The short-dashed line in Fig.~\ref{fig:hrd} shows the instability 
boundary of low-order radial and nonradial p-mode oscillations. 
In the long nearly vertical region of $4.6 \ga \log T_{\rm eff} \ga 4.3$,
the kappa-mechanism at the Fe opacity bump excites low-order modes,
which correspond to $\beta$ Cep variables
\citep{ki92,mo92,pa99}.

The instability boundary bends and becomes horizontal due to
strange mode instability \citep{ki93,gl94,sa98} which occurs
when the luminosity to mass ratio is sufficiently high 
($L/M \ga 10^4 L_\odot/M_\odot$) and radiation pressure is dominant
at least locally in the envelope.
The instability boundary is essentially the same as that obtained
by \citet{ki93} for $Z=0.02$.
In the hottest part, the strange modes are associated with the Fe opacity
peak, while in relatively cooler part the contribution from the 
opacity peak at the second helium ionization becomes important.

The vertical part around $\log T_{\rm eff}\sim3.8$ is the well known
blue edge of the Cepheid instability strip. 

A narrow vertical region indicated by a long-dashed line 
around $\log T_{\rm eff}\sim 3.8 - 3.9$ in Fig.~\ref{fig:hrd} indicates
the excitation of relatively high order radial and nonradial modes
due the strange mode instability associated with the hydrogen ionization zone.
We will discuss these modes in \S\ref{sec:high}.

Monotonously unstable modes exist above the dotted line in the most 
luminous part of Fig.~\ref{fig:hrd}, which will be discussed  
in \S\ref{sec:secular}. 

In the regions surrounded by blue and red solid lines, nonradial
g-modes are excited by the Fe opacity bump; red and blue lines for 
$\ell=1$ and $\ell=2$ modes, respectively.
Those g-modes can be excited even in the post-main-sequence models
if the mass exceeds $12M_\odot$, because a convection zone
above the hydrogen burning shell can block some g-modes from 
penetrating to the core.

\subsection{Oscillatory convection modes}\label{sec:conv}

\begin{figure*}
\epsfig{file=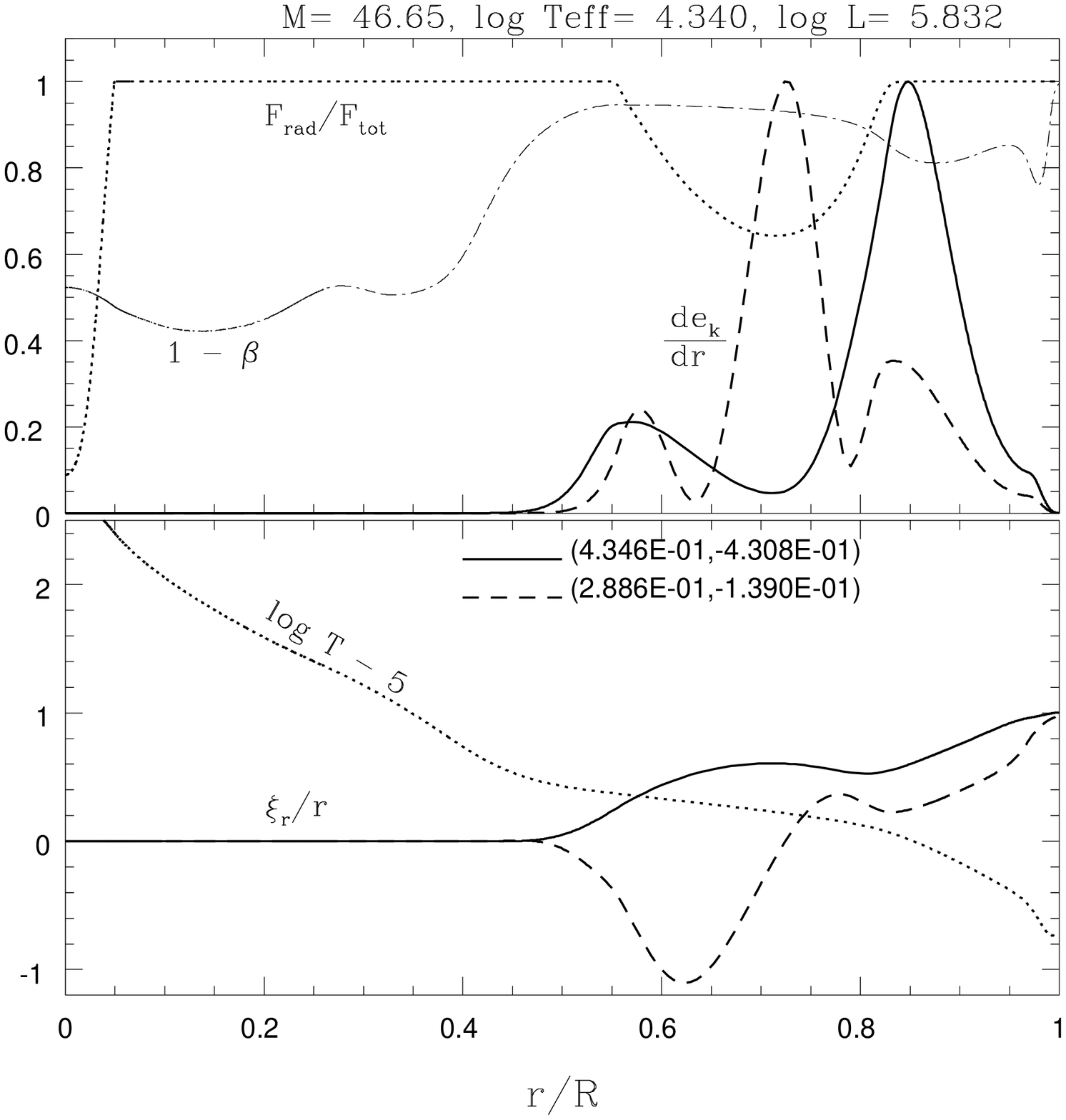,width = 0.49\textwidth}
\epsfig{file=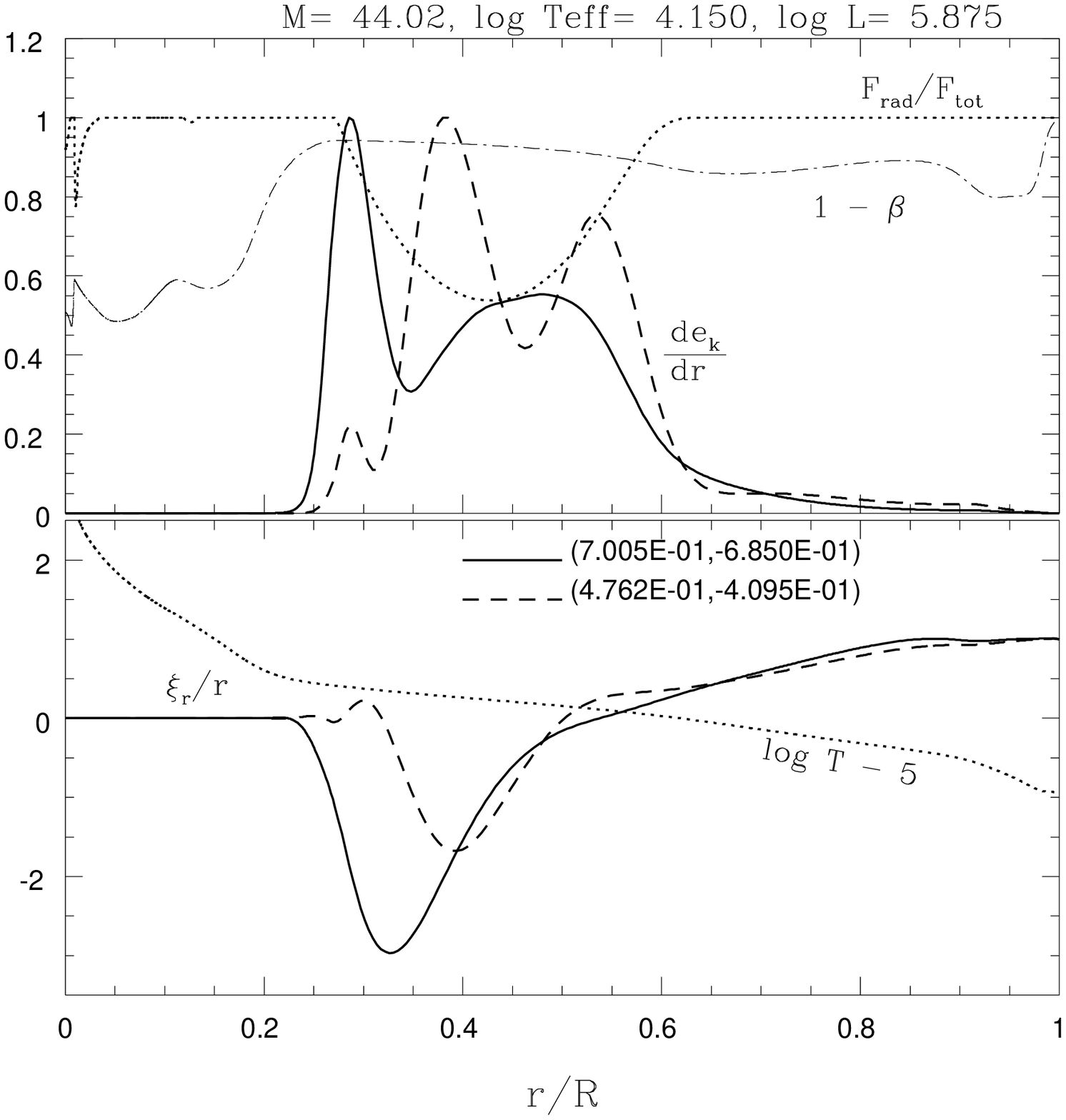,width = 0.49\textwidth}
\caption{Runs of radial displacements (bottom panels) and kinetic 
energy distributions (top panels) as function of fractional radius
($r/R$) for selected oscillatory convection modes
of $\ell=1$ in two supergiant models with different effective temperatures
but with the same initial mass of $60M_\odot$.
Also shown are runs of temperature and the ratios of radiation to total flux 
($F_{\rm rad}/F_{\rm tot}$) and radiation pressure to total pressure ($1-\beta$).
The numbers in parentheses are the real and imaginary parts of 
eigenfrequencies normalized by $\sqrt{GM/R^3}$.
}
\label{conv_amp}
\end{figure*}

The dash-dotted lines in Fig.~\ref{fig:hrd} are the ranges where  
oscillatory convection modes would be observable; red and blue lines 
are for $\ell=1$ and $\ell=2$ modes, respectively.
The observability is determined based on the ratio
between the photospheric amplitude and the maximum amplitude
in the interior, which is discussed below.

Fig.~\ref{conv_amp} shows distributions of radial displacements 
($\xi_r/r$, bottom panels) and
kinetic energy density ($de_k/dr$, top panels) of oscillatory convection modes
of $\ell=1$ in two (hot and relatively cool) models from 
the $M_i=60M_\odot$ ($Z=0.02$) evolutionary sequence. 
Two convection modes for each model are shown with selected physical 
quantities. 
A convection zone can be recognized as a zone where the ratio of 
radiative to total energy flux is less than unity;i.e.,
$F_{\rm rad}/F_{\rm tot} < 1$.
The Fe-convection zone occurring in a temperature range
of $5.4 \ga \log T \ga 4.9$ has a considerable thickness but contains
very small mass ($\la 10^{-3}M_\odot$ for the hotter model, and
$\sim 4\times10^{-3}M_\odot$ for the cooler model).
Because the gas density is very low in the envelope of a massive star, 
the convective flux  is less than 50\% of the total energy flux. 
In addition, the low density makes the gas pressure much smaller than
the radiation pressure; i.e., $\beta \ll 1$ with $\beta$ being 
the ratio of gas to total pressure.

\begin{figure}
\epsfig{file=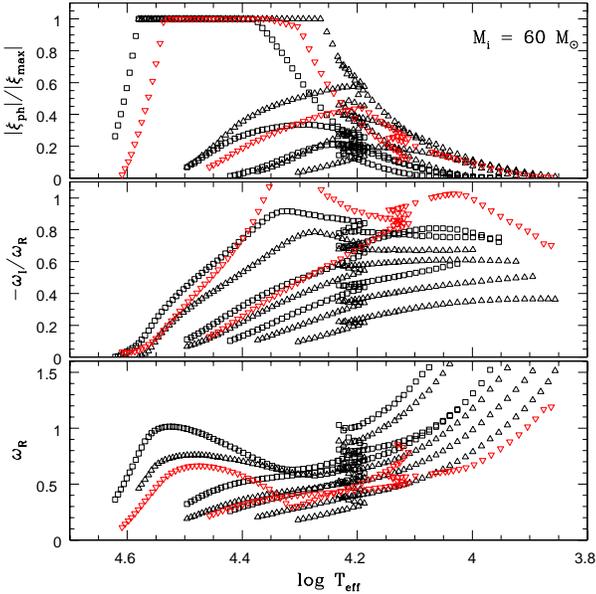,width=0.49\textwidth}
\caption{Some properties of oscillatory convection modes versus 
effective temperature along the evolutionary track of $M_i=60M_\odot$
with $Z=0.02$; bottom panel: Real part of normalized frequencies;
middle panel: growth rate; top panel:the ratio of the amplitude of the
photosphere to the maximum amplitude. Triangles and squares 
are for $\ell=1$ and $\ell=2$, respectively.
Red inverted triangles are for $\ell=1$ modes in models calculated
with {\it a mixing-length of one pressure scale-height}.
}
\label{conv_te}
\end{figure}

Two modes shown for each model in Fig.~\ref{conv_amp} are the two shortest
period convection modes that are most likely observable; i.e., 
ratios of the photospheric amplitude to the maximum amplitude in the
interior are largest (see also Fig.~\ref{conv_te}).
The numbers in parentheses in the bottom panels of Fig.~\ref{conv_amp}
are normalized (by $\sqrt{GM/R^3}$) complex eigenfrequencies;
the real part represents oscillation frequency, and imaginary
part growth or damping rate. When the imaginary part is {\it negative}
the oscillation grows. 
The eigenfrequencies shown in Fig.~\ref{conv_amp} indicate that
the growth times are comparable to the periods. The periods are comparable
to the periods of g-modes.

The kinetic energy of the oscillatory convection modes is confined
to and slightly above the convection zone in the hotter model, while 
it is well confined to the convection zone in the cooler model. 
For the convection modes shown in this figure, 
the oscillation amplitude at the stellar surface is comparable to
that in the convective zone, indicating those modes are 
very likely observable.   

Since oscillatory convection modes have short growth-times, we expect that
they have substantial amplitudes in the convection zone.
For such a mode to be observable, however, the amplitude at the stellar
surface should be, at least, considerable relative to the amplitude
in the convection zone.
We assume that the possibility of detecting 
these convection modes can be measured by the 
ratio of the photospheric amplitude to the maximum amplitude in the
interior. 
The ratio differs for different modes in a model and 
changes as the stellar parameters change.
The top panel of Fig.~\ref{conv_te} shows variations of the ratio
as a function of effective temperature along the evolutionary
track of $M_i=60M_\odot$ for a few convection modes of 
$\ell=1$ (triangles) and $\ell=2$ (squares).
This figure shows that although many convection modes exist
in a model, only one or two highest-frequency modes (for each $\ell$) 
have large values of
the amplitude ratio and hence are potentially observable.
Generally, the amplitude ratio of each mode has a broad peak 
as a function of effective temperature, and the peak value
tends to be larger for a mode with larger frequency (bottom panel) and 
larger growth rate (middle panel). 
In other words, the visibility is highest for largest frequency modes with
highest growth rates in a certain effective temperature range
($4.2 \la \log T_{\rm eff} \la 4.6$  for the $M_i=60M_\odot$ models).

\begin{figure}
\epsfig{file=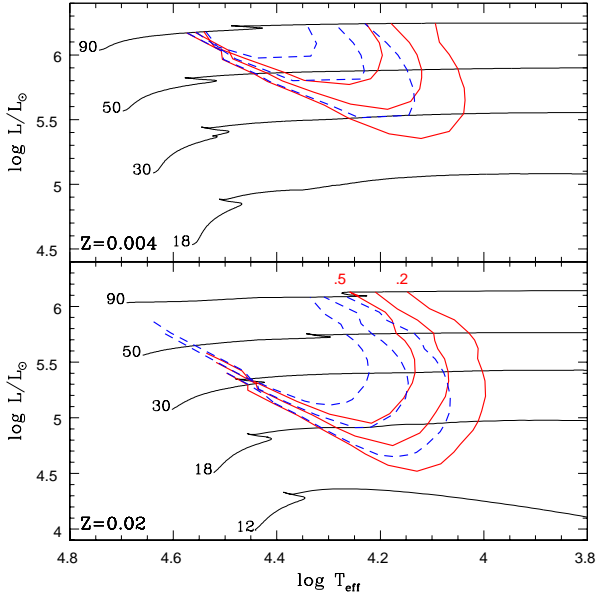,width=0.49\textwidth}
\caption{Contours of $0.5, 1/3$ and $0.2$ for the ratio of the amplitude 
at the photosphere to the maximum amplitude in the interior.
Solid and dash-dotted lines are for $\ell=1$ and $\ell=2$ modes,
respectively. 
The bottom panel is for
normal metallicity of $Z=0.02$ and the top panel for a metal-poor
composition of $Z=0.004$.
}
\label{conv_cont}
\end{figure}

Fig.~\ref{conv_cont} shows contours of the amplitude ratio for the most
visible convection modes for $\ell=1$ (solid lines) 
and $\ell=2$ (dashed lines).
Generally, the visibility of the oscillatory convection modes is better
in more luminous and hotter models.
We assume in this paper that
these modes would be well visible when the ratio exceeds 
$0.2$. The contours for the ratio of $0.2$ 
are shown in Fig.~\ref{fig:hrd}.

The top panel of Fig.~\ref{conv_cont} shows the contours obtained for
models with $(X,Z) = (0.716,0.004)$.
Since the opacity is much lower than that for the
case of $Z=0.02$, the convection zone around the Fe opacity peak is
less extensive.
Nonetheless, observable oscillatory convection modes with low degree $\ell=1, 2$
persist, although the contour for a
value of the amplitude ratio shifts upward by $\Delta \log L \approx 0.8$
as compared to the case of the standard composition.
This indicates that oscillatory convection modes can cause observable
variations even in the SMC if the stars are luminous enough. 

\begin{figure}
\epsfig{file=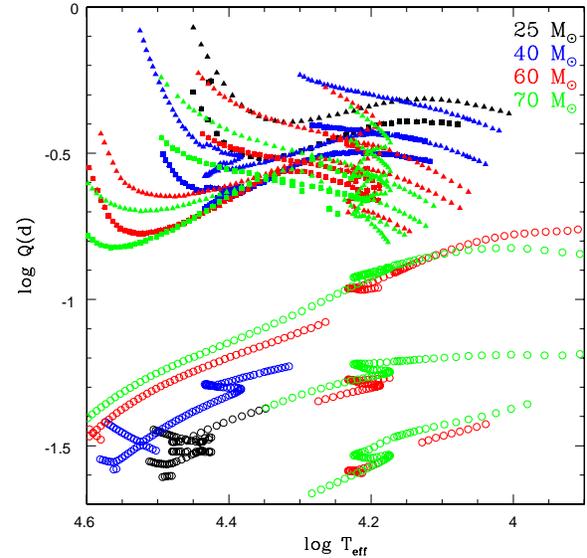,width=0.49\textwidth}
\caption{Q-values of convection modes (filled triangles for $\ell=1$,
filled squares for $\ell=2$) and excited radial
modes (open circles) versus effective temperatures along the
evolutionary tracks of $M_i = 25, 40, 60$ and  $70 M_\odot$.
}
\label{conv_Q}
\end{figure}

The growth rates of the oscillatory convection modes 
(middle panel, Fig.\ref{conv_te}) are very high,
especially in relatively cooler models;
they tend to decrease as $T_{\rm eff}$ increases.
The reason is probably that in hotter models kinetic energy distribution
shifts outward and hence a large fraction of the kinetic energy 
resides above the convective zone boundary losing the driving.  

It has been thought that the pulsation modes with the highest growth rates 
are associated with strange modes \citep[eg.][]{gg90b,sa98}. 
The growth rates of the convection modes are even larger 
than those of radial strange modes in massive stars \citep{gl93,ki93}.
The oscillation frequencies of convection modes 
lie in the g-mode range as seen in
the bottom panel of Fig.~\ref{conv_te}, indicating the periods of
convection modes are longer than radial modes.

Fig.~\ref{conv_Q} shows Q-values of convection modes with the amplitude
ratio larger than $0.2$ (filled symbols)
compared with excited radial modes (open circles) for some selected mass models,
where the Q-value is defined as
$\Pi\sqrt{\overline\rho/\overline\rho_\odot}$ with
$\Pi$ and $\overline{\rho}$ being, respectively, pulsation period and 
mean density \citep[see e.g.][]{cox80}.
The Q-value of a convection mode decreases rapidly
as the effective temperature decreases in the hottest and  the coolest
$T_{\rm eff}$ ranges, while it changes little in the intermediate
range of effective temperature.
At a given effective temperature, Q-values of oscillatory convection modes
tend to be smaller in more massive models.

In very massive ($M_i=60, 70M_\odot$) models, radial pulsations are excited
even in cooler ($\log T_{\rm eff} \la 4.2$) models due to mainly the strange
mode effect that works around the second He ionization zone \citep{gl93,ki93}.
The Q-values of these strange modes increases as the effective temperature
decreases; i.e., as the depth of the He II ionization zone increases. 
Fig.~\ref{conv_Q} indicates that periods of oscillatory convection modes
are much longer than those of radial pulsations in hotter models, although
in cooler and massive models the differences are not very large.
\citet{Lam98} found that the periods of microvariations of LBVs
are orders of magnitudes longer than those predicted for strange modes
by \citet{ki93}.
This indicates that oscillatory convection modes would be responsible 
for long-period microvariations in LBVs (see \S\ref{obscomp} below).

Red inverted triangles in Fig.~\ref{conv_te} present the property of
oscillatory convection modes of $\ell=1$ in models calculated
with a mixing-length of {\it one pressure scale-height} rather than 
$1.5$ pressure scale-hight adopted in standard models.
A smaller mixing-length makes the super-adiabatic temperature gradient 
in the Fe-convection zone larger.
In the models with the smaller mixing-length,
real parts of eigenfrequencies of oscillatory convection modes tend
to be smaller, and the amplitude tends to be slightly more confined to
the convection zone. 
Fortunately, the qualitative property of the 
oscillatory convection modes
is insensitive to the mixing-length parameter.

\subsection{Monotonously unstable mode} \label{sec:secular}

\begin{figure}
\epsfig{file=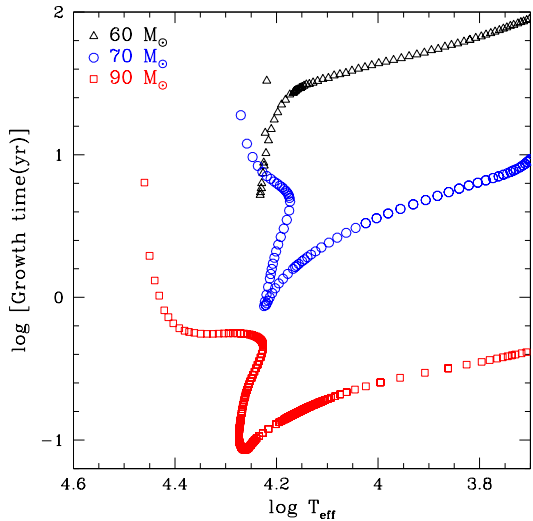,width=0.45\textwidth}
\caption{
Growth-time versus effective temperature for monotonously unstable
modes in models with $M_i=60M_\odot, 70M_\odot$ and $90M_\odot$.
}
\label{growth_sec}
\end{figure}

In very massive models ($M_i\ge 60M_\odot$) a monotonously unstable 
radial mode (with a purely imaginary eigenfrequency)
is found in the range on the HR diagram indicated by a 
dotted line in Fig.~\ref{fig:hrd}.
Fig.~\ref{growth_sec} shows growth time as a function of 
effective temperature for those unstable modes in models with
different initial masses.
The growth times tend to be shorter for more massive stars, ranging from
a month to a hundred years, which are much faster than the evolutionary 
change.
It is interesting to note that the range of the growth times includes
the timescales of S Dor (SD)-phase \citep{vanG01}.
To the author's knowledge, the presence of monotonously
unstable modes in very massive stars was not known before.
The presence 
was unrecognized in 
the non-linear radial pulsation analyses performed before by various authors,
probably  because the monotonously unstable  mode was 
concealed by the more fast growing
strange mode pulsations.

\begin{figure}
\epsfig{file=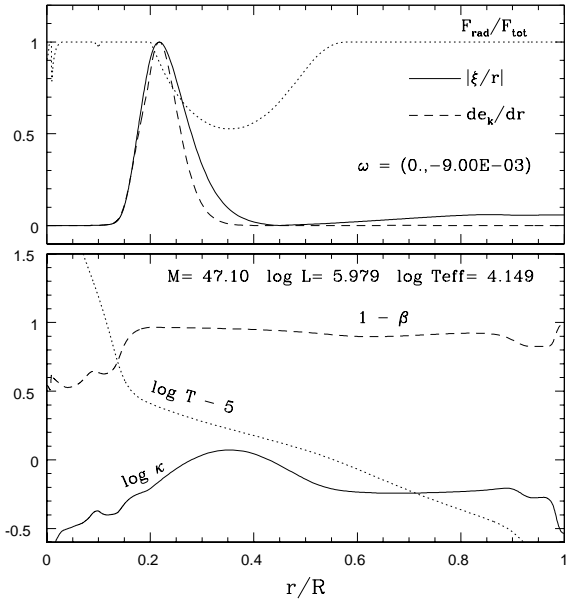,width=0.49\textwidth}
\caption{
The amplitude of displacement ($|\xi/r|$) and 
kinetic energy distribution ($de_k/dr$) and selected physical quantities
as functions of the fractional radius ($r/R$) for a model with
$M_i= 70M_\odot$, where $\kappa$ is opacity, and other quantities
are the same as in Fig.~\ref{conv_amp}.
}
\label{fig:secmode}
\end{figure}

Amplitude and kinetic energy distribution of the monotonously unstable mode
in a model with $M_i=70M_\odot$ are shown in Fig.~\ref{fig:secmode}.
The amplitude and kinetic energy  have a peak around the bottom 
of the convective zone
generated by the Fe-opacity peak; in this model the convective zone
ranges from $r/R=0.20$ to $0.595$.
Although, the kinetic energy, as well as amplitude, is confined around 
the bottom of the convection zone, the amplitude, after attaining a
minimum at $r/R \approx 0.45$, gradually increases
toward the surface above the convection zone.
In a substantial range in this model
the ratio of the gas to radiation pressure is very small (i.e., $\beta \ll 1$).
Then,
\begin{equation}
{L_{\rm rad}\over L_{\rm Edd}} \approx (1-\beta) \approx 1,
\end{equation}
where $L_{\rm rad}$ is radiative luminosity and 
$L_{\rm Edd}$ the local Eddington luminosity defined as 
\begin{equation}
L_{\rm Edd}\equiv 4\pi cGM_r/\kappa
\end{equation}
with $c$ being the velocity of light and $M_r$ the mass within 
the sphere of radius $r$.
This suggests that the monotonously unstable mode arises because
the radiative luminosity is very close to the local Eddington luminosity
in a substantial range of the envelope.

\begin{figure}
\epsfig{file=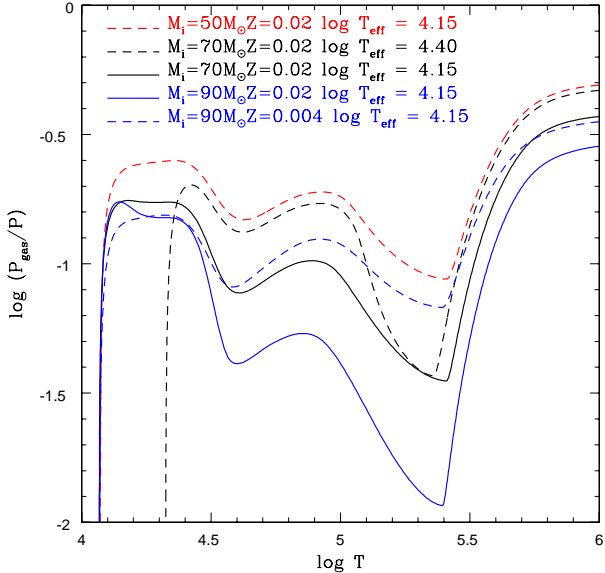,width=0.49\textwidth}
\caption{
Distribution of the ratio $P_{\rm gas}/P$ ($\equiv\beta$) in the interior
of selected models.
Solid lines are used for the models which has  a monotonously 
unstable mode, while broken lines for the models having none.
}
\label{fig:beta}
\end{figure}

Fig.~\ref{fig:beta} shows the distribution of $\beta$ in selected models,
where solid lines are used for models which have a monotonously unstable mode, 
while dashed lines are for models without it.
There are two minima of $\beta$ at $\log T\approx 5.4$ (Fe opacity peak) and
at $\log T\approx 4.6$ (He II ionization). 
Apparently, $\beta$ must be well below 0.1 in a range including the both opacity
peaks for a monotonously unstable mode to appear.
Note that the hotter model of $70M_\odot$ (black dashed line) has a
distribution of $\beta$ around $5.2\la \log T \la 5.4$ similar to
that of the cooler model (black solid line), but $\beta$ in the range $4.5\la \log T \la 5$
is significantly higher than that in the cooler model, which inhibits a monotonously
unstable mode in the hotter model.
With a low-metal abundance of $Z=0.004$ no monotonously unstable mode is found
even in models of $M_i=90M_\odot$. 
The reason is obvious from the blue dashed line in Fig.~\ref{fig:beta}; 
$\beta$ at the Fe opacity peak is substantially larger, although a further
increase in mass seems to cause monotonously unstable modes.

The presence of such a monotonously unstable  mode probably corresponds 
to the presence 
of an optically thick wind as discussed by \citet{ka92} and \citet{nu02} for
WR stars.
It is known that optically thick winds also occur in nova models. 
\citet{ka09} have shown that for non-extreme nova cases
both static models and models with winds are possible.
The presence of a monotonously unstable mode
in a massive evolutionary model might indicate that the static model
transits to a model with an optically thick wind.

It is interesting to note that the boundary in the HR diagram
for the presence of monotonously unstable  modes 
(dotted line in Fig.~\ref{fig:hrd})
roughly coincides with the  Humphreys-Davidson (HD) limit \citep{hd79}; 
this may suggest the HD limit to be related to 
the presence of optically thick winds. 

\subsection{High-oder modes excited at H ionization zone}  \label{sec:high}
In the narrow vertical region around $\log T_{\rm eff}\sim3.9$ seen
in Fig.~\ref{fig:hrd}, relatively high order ($n\sim 6 - 10$) radial as well as
nonradial modes are excited at the
hydrogen ionization zone at $\log T \approx 4$.
Since the normalized frequency of the excited mode increases as
the effective temperature decreases, the red-edge is formed 
where the frequency exceeds the critical  frequency 
given as ${1\over2}\sqrt{\Gamma_1 R/H_p}$ with the
pressure scale height $H_p$ and $\Gamma_1\equiv (d\ln P/d\ln\rho)_{\rm ad}$.
(If the frequency exceeds the critical frequency,
the pulsation is not reflected at the outer
boundary and is expected to be dissipated in the outermost layers.)

\begin{figure}
\epsfig{file=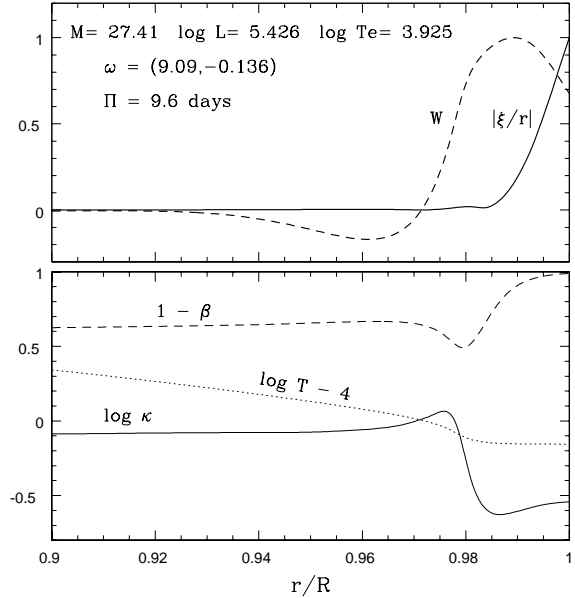,width=0.49\textwidth}
\caption{
Amplitude distribution and work curve (top panel) for an excited
9th-order radial pulsation. Some physical quantities are
shown in the bottom panel. 
}
\label{radhigh}
\end{figure}

Fig.~\ref{radhigh} shows the amplitude distribution and the work curve for
a high-order radial mode excited in a model relatively close
to the blue edge. 
Obviously the excitation occurs at $\log T \approx 4$, in the hydrogen 
ionization zone.
The amplitude of the mode is extremely confined to the surface layers,
where the radiation pressure is much larger than the gas pressure.
Since unstable modes of this type are present even in the 
non-adiabatic reversible
\citep[NAR;][]{gg90b} approximation 
where thermal-time is artificially set to be zero,
we can identify such a mode as a strange mode trapped above 
the opacity peak of the hydrogen ionization.
A similar mode with a similar frequency is excited for $\ell=1$ and $\ell=2$.
These excited nonradial modes probably correspond to the trapped 
modes discussed in \citet{afg09}.
In addition to the trapped modes \citet{afg09} found several untrapped 
modes to be excited around the trapped mode frequency,
by using the Riccati-shooting method discussed in \citet{gg90a}.
The present study using a finite difference method, however, 
could not find those untrapped modes.
According to \citet{afg09} those modes have many nodes with considerable amplitude
in the deep interior around $r/R\sim10^{-2}$.
The finite difference code even with  $\sim10^4$ grid points seems ineffective
to find such untrapped modes.

\section{Comparison with observed periods}\label{obscomp}

\begin{figure}
\epsfig{file=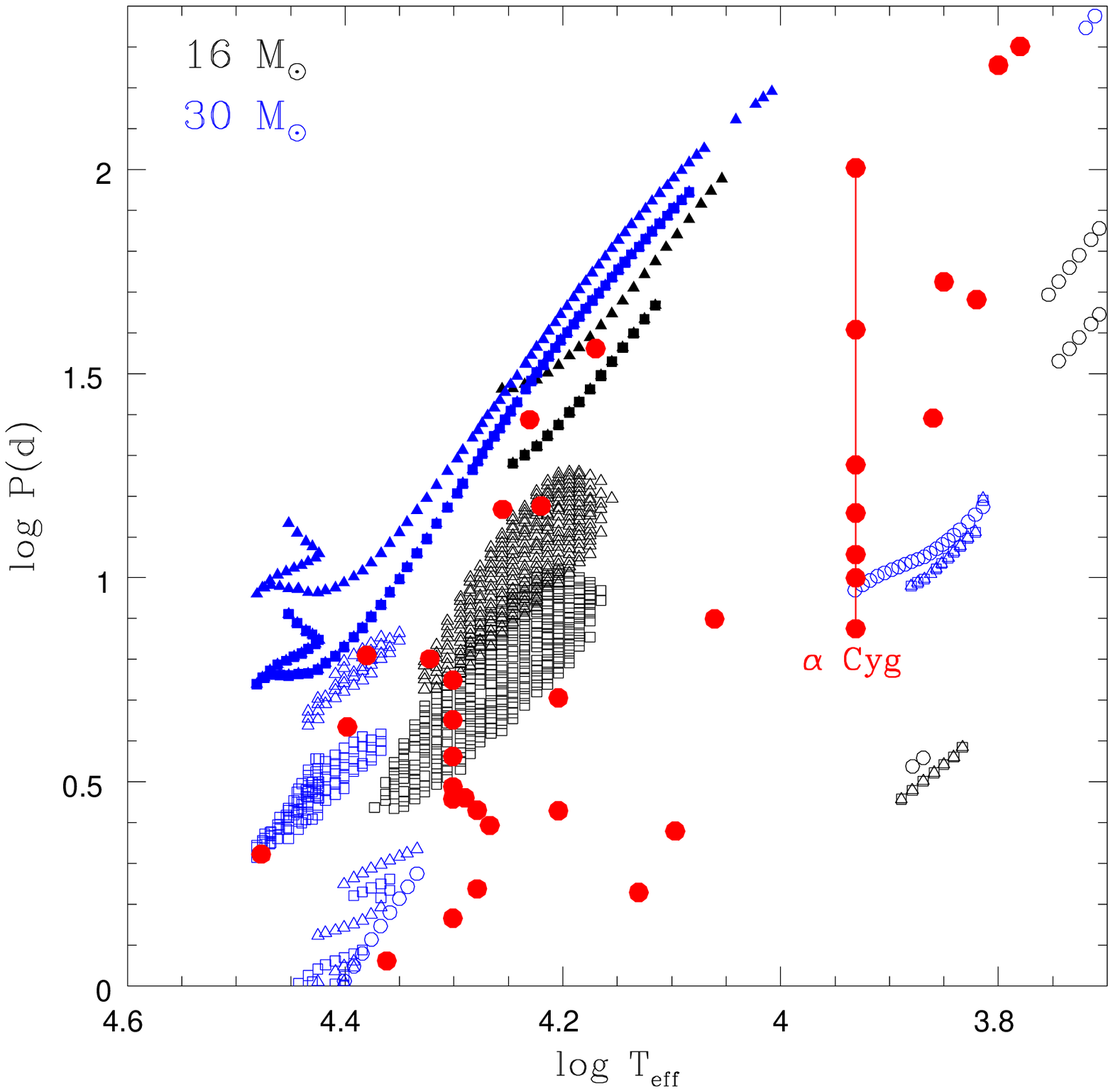,width=0.49\textwidth}
\caption{
Periods of radial and nonradial modes excited in $16$ and  $30M_\odot$ models
are compared with observed periods of relatively less luminous 
($\log L/L_\odot \le 5.5$) supergiants (large dots) which consists of
periodic B stars from \citet{Lef07} and relatively cooler less luminous
$\alpha$ Cyg variables from \citet{vanL98} and \citet{vanG02}.
The periods of $\alpha$ Cyg itself are adopted from \citet{lu76},
while the effective temperature from \citet{sc08}.
Triangles and squares correspond to $\ell=1$ and $\ell=2$,
respectively. 
Filled triangle and squares are convection modes for which the amplitude
ratios (see \S\ref{sec:conv}) are larger than 0.2. 
Open circles are radial modes.
Swarms of $\ell=1, 2$ modes in a range of $4.35 \ga \log T_{\rm eff} \ga 4.15$
for $16M_\odot$ and in a range of $4.5 \ga \log T_{\rm eff} \ga 4.35$
for $30M_\odot$ are high-order g-modes (i.e., SPBsg-type oscillations).}
\label{te_peri_obs16}
\end{figure}

Fig.~\ref{te_peri_obs16} compares periods of 
radial and nonradial modes excited in  $16M_\odot$ and $30M_\odot$ models
with observed periods of relatively less luminous ($\log L/L_\odot < 5.5$)
supergiant variables as function of $T_{\rm eff}$.
Those variables consist of most of the periodic B stars analyzed by
\citet{Lef07}, and relatively cooler $\alpha$-Cygni variables including
$\alpha$ Cyg itself. 
Most of the periods of the hotter ($\log T_{\rm eff}\ga 4.15$) less luminous
supergiants seem consistent with oscillatory convection modes, 
SPB-type g-modes (SPBsg), or p-modes.   

Relatively cool ($\log T_{\rm eff} \la 3.95$) stars in Fig.~\ref{te_peri_obs16}
are less luminous $\alpha$-Cygni variables.
They lie in the $T_{\rm eff}$ range where high-order p-modes are excited
(Fig.~\ref{fig:hrd}), although most of the periods are longer than those of 
high-order p-modes.
Probably they are pulsating in g-modes excited in a similar range of
$T_{\rm eff}$ as found by \citet{afg09}, who found that g-modes
having periods of ranging from about 10 to 20 days are excited
at the hydrogen ionization zone, although 
a large fraction of their kinetic energy resides in the deep interior.

No modes can be assigned to the three stars in a range of 
$4.15 \ga \log T_{\rm eff} \ga 4.05$ in
Fig.~\ref{te_peri_obs16}, which are the coolest members in the data from
\citet{Lef07} (Fig.\ref{fig:hrd}).
The cause of the oscillations in these stars is not clear.

\begin{figure}
\epsfig{file=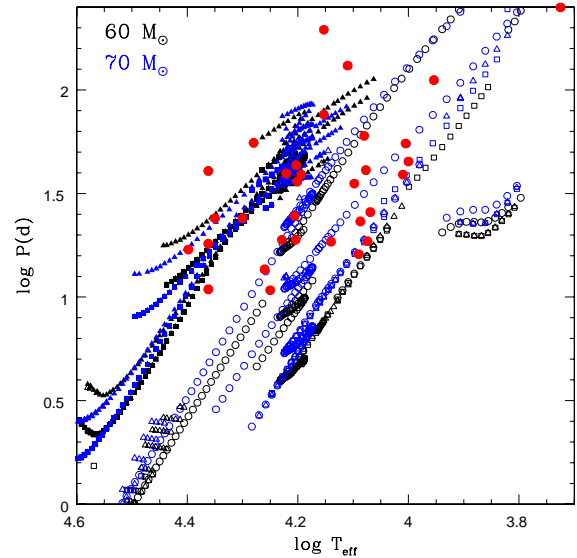,width=0.49\textwidth}
\caption{
Periods of radial and nonradial modes excited in
high mass models of $M_i=60M_\odot$ and $70M_\odot$ with $Z=0.02$
are compared with observed periods of luminous ($\log L/L_\odot \ge 5.5$)
supergiants (red dots) which consists of microvariations of LBVs 
from \citet{Lam98} and luminous $\alpha$ Cyg variables from \citet{vanL98} and 
\citet{vanG02}.
Symbols for theoretical periods have the same meanings as in
Fig.~\ref{te_peri_obs16}.
}
\label{te_peri_obs60}
\end{figure}

Fig.~\ref{te_peri_obs60} compares periods of radial and nonradial modes
excited in higher mass models of $M_i=60$ and  $70M_\odot$ with observed ones
for microvariations in luminous ($\log L/L \ge 5.5$) supergiants
which consist of LBVs and luminous $\alpha$ Cyg variables.
In these massive stars, radial and nonradial strange modes 
are excited (open symbols), and oscillatory convection modes 
(filled triangles and squares) are likely observable (the amplitude ratio is
larger than 0.2) 
in a wide range of $T_{\rm eff}$,  while SPB-type g-modes are not excited.

Fig.~\ref{te_peri_obs60} indicates that most of the periods 
of microvariations of LBVs and luminous $\alpha$ Cyg variables
are consistent with the periods of oscillatory convection modes
and strange modes.

The comparisons between observed and theoretical periods 
in Figs.~\ref{te_peri_obs16} and ~\ref{te_peri_obs60}
indicate that $\alpha$-Cygni variables are 
inhomogeneous. Luminous ones are similar to the microvariations
of LBVs which are identified as oscillatory convection modes or strange modes,
while less luminous cooler ones, including $\alpha$ Cyg itself,
seem to be g-modes (and possibly including high-order strange modes) excited
at the hydrogen ionization zone.

\section{Low metallicity cases}

\begin{figure}
\epsfig{file=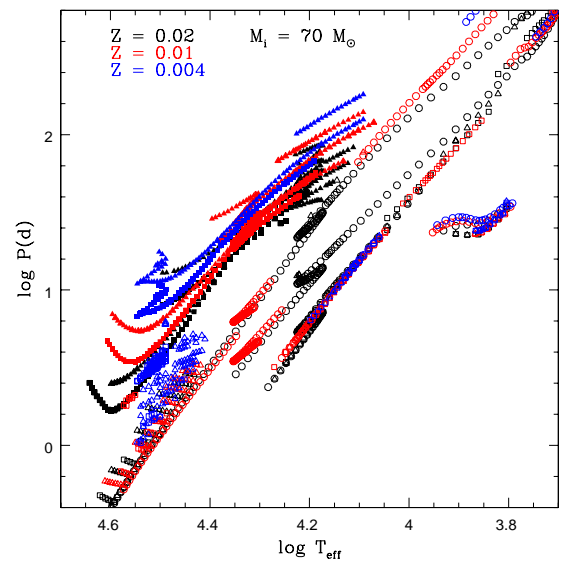,width=0.49\textwidth}
\caption{
Periods of radial and nonradial oscillations excited in models
with different heavy element abundances are compared.  
The symbols for the excited modes are the same as in Fig.~\ref{te_peri_obs16}.
}
\label{te_peri_compz}
\end{figure}

Since there are some LBVs and $\alpha$ Cyg variables in SMC,
it would be useful to discuss the results for low metal abundances. 
Fig.~\ref{te_peri_compz} compares periods of radial and nonradial
modes excited in models with $M_i=70M_\odot$ having different
metal abundances.
The difference between the cases of $Z=0.02$ and $Z=0.01$ is small,
while the case of $Z=0.004$ is significantly different from the former cases.
Because of the Fe-opacity peak being reduced, strange mode instability
for low-order radial and nonradial modes are reduced significantly
for the $Z=0.004$ case, while the excitation of SPBsg-type g-modes 
seems to be enhanced in the lowest metallicity case 
(the reason is not clear).

It is interesting to note that even for the low metallicity of $Z=0.004$
oscillatory convection modes still seem to be observable in a considerable
range of $T_{\rm eff}$ (see also Fig.~\ref{conv_cont}).

\begin{figure}
\epsfig{file=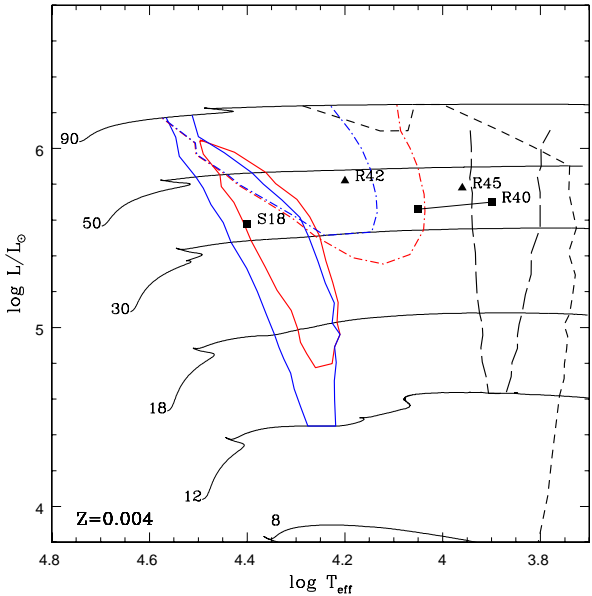,width=0.49\textwidth}
\caption{The same as Fig.~\ref{fig:hrd} but for a metal-poor 
composition of $(X,Z)=(0.716,004)$.
Also plotted are some S Dor variables \citep{vanG01} and 
$\alpha$ Cyg variables \citep{vanL98} in SMC. 
}
\label{fig:hrdz004}
\end{figure}

Fig.~\ref{fig:hrdz004} is the same as Fig.~\ref{fig:hrd} but for 
$Z=0.004$; i.e., shows instability ranges
for radial and nonradial modes and observable ranges for oscillatory
convection modes in the HR diagram.
Compared to the case of $Z=0.02$ (Fig.~\ref{fig:hrd}), the instability
region for low-order radial and nonradial p-modes 
(i.e., $\beta$ Cep instability strip)
has disappeared completely. The instability boundary for 
the low-order strange mode instability has shifted upward considerably, 
which agrees with the result of \citet{ki93}.

\begin{figure}
\epsfig{file=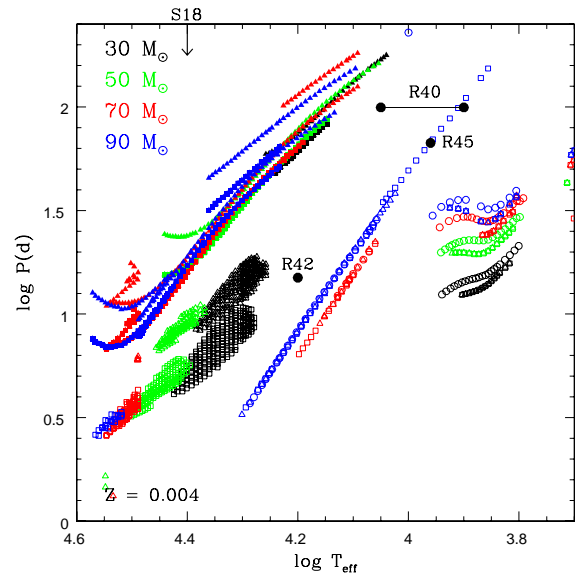,width=0.49\textwidth}
\caption{
Periods of excited modes in metal-poor models ($Z=0.004$) of 
$M_i= 30, 50, 70$, and $90M_\odot$ as function of $T_{\rm eff}$ are compared with
observed periods of microvariations of LBVs  
and $\alpha$ Cyg variables in SMC. 
Symbols for theoretical periods have the same meanings as in
Fig.~\ref{te_peri_obs16}.
}
\label{te_peri_smc}
\end{figure}

On the other hand, less affected are the SPBsg-type g-mode instability regions  
and the observable region of oscillatory convection modes.
The instability region for high-order modes in $3.95\ga \log T_{\rm eff} \ga 3.8$
is not affected because they are excited at hydrogen ionization zone.
Positions of some LBVs (S Dor variables) and $\alpha$ Cyg variables
in SMC are also plotted in Fig.~\ref{fig:hrdz004}.
Fig.~\ref{te_peri_smc} compares quasi-periods of the microvariations of 
those stars with theoretical periods predicted for
metal poor models with initial masses ranging from  $M_i = 30$ 
to $90M_\odot$.  

In contrast to the Galactic and LMC cases, oscillatory convection
modes seem to be responsible for none of the SMC stars shown 
in these figures. 
Fig.~\ref{te_peri_smc} seems to indicate 
that periods of low-order strange modes
are comparable with those of the microvariations of 
R40 and R45 (and probably R42).
For the strange mode interpretation to be true, the
initial mass must be greater than $\sim 80M_\odot$.
Fig.~\ref{fig:hrdz004}, however, indicates the initial mass of
these stars to be $40 - 50M_\odot$; i.e., 
the luminosities of these stars seem too low
at least by a factor of two.
Therefore, the mode identification for the microvariations of R40, R42 and R45
is unclear.

For S18, an $\alpha$ Cyg variable in the SMC, only the effective
temperature is indicated in Fig.~\ref{te_peri_smc} because
the quasi-periods of its variability are uncertain.
\citet{vanG02} found three types of variations on timescales of
a few years, $\sim 150$ days, and a few days.
Figs.~\ref{fig:hrdz004} and \ref{te_peri_smc} indicate the shortest timescale 
variation to correspond to the SPBsg type g-modes. 
For the longer-periods variations, however, no excited oscillation modes 
can be assigned.

\section{Conclusions}
The stability of radial and nonradial modes in massive stars
($90 \ge M_i/M_\odot \ge 8$) was investigated.
The $\kappa$-mechanism associated with the Fe-opacity bump 
at $T\sim2\times10^5$K excites low-order radial and nonradial p-modes
in main-sequence and post-main-sequence models ($\beta$ Cep instability strip)
and low-degree  high radial order g-modes in post-main sequence
models (supergiant SPB stars; SPBsg ).
Interestingly, it is found that 
under a low-metal composition of $Z=0.004$, the SPBsg instability
region still remains, while the $\beta$ Cep instability strip 
disappears.  

In luminous models with $L/M \ga 10^4L_\odot/M_\odot$ (for $Z=0.02)$
strange mode instability occurs associated with opacity peaks due to
Fe and He II ionizations. 
For the metal-poor composition $Z=0.004$ strange mode instability
appears only in more luminous models.

In a narrow strip of $3.95 \ga \log T_{\rm eff} \ga 3.85$, high-order
radial and nonradial pulsations are excited at the hydrogen-ionization
zone. The pulsation amplitude is strongly confined to the outermost
layers.

Low-$\ell$ oscillatory convection modes 
associated with the Fe convection zones in massive stars are found. 
They are likely observable in sufficiently luminous models in a considerable range
of effective temperature.
The oscillatory convection modes 
tend to have growth times comparable with
the periods. Their periods are longer than those of strange modes 
or SPBsg-type g-modes, and are comparable to
those of long-period microvariations of LBVs and 
luminous $\alpha$ Cyg variables.
Rapid growth rates of the oscillatory convection modes 
are consistent with the presence of irregularities 
in the microvariations.

In addition to the oscillation modes, a monotonously 
unstable mode is found
in luminous models with $\log L \ga 5.9$ and 
$\log T_{\rm eff} \la 4.3$.
The boundary in the HR diagram for the presence of the monotonously unstable
mode roughly coincides with the HD limit.  
The growth rate is more than hundred 
times faster than the stellar evolution rate.
It probably suggests the presence of an optically thick wind in such a 
luminous star.  

Finally, it should be noted that in the present stability analysis
the effect of stellar winds is not 
included in the outer boundary condition. 
In the future, we have to clarify how the stellar wind
affects the stability of oscillations of massive stars.

\section*{Acknowledgments}
I am grateful to Alfred Gautschy for making sample calculations with
an alternative numerical approach to
confirm the presence of overstable low-degree convection modes,
and for helpful comments on a draft of this paper.
I am also grateful to the anonymous referee for
useful comments, which have improved the paper 
considerably.

%\bsp

\label{lastpage}

\end{document}